\documentclass{optica-article}

\journal{opticajournal} 

\articletype{Research Article}

\usepackage{lineno}
\usepackage{listings}
\usepackage{braket}
\usepackage{float}
\usepackage{graphicx}
\usepackage[version=4]{mhchem}
\usepackage{booktabs}
\usepackage{array}
\usepackage{setspace} 
\usepackage{amsmath}
\usepackage{esint}

\usepackage{amssymb}

\newcommand{\E}{\mathbb{E}}
\newcommand{\Tr}{\operatorname{tr}}

\newcommand{\vect}[1]{\mathbf{#1}}      
\newcommand{\mat}[1]{\mathbf{#1}}       

\newcommand{\eqnref}[1]{Eq.~\eqref{#1}}
\newcommand{\secref}[1]{Sec.~\ref{#1}}
\newcommand{\figref}[1]{Fig.~\ref{#1}}

\begin{document}

\title{Topology-optimized distributed 3d anisotropic Raman emission}
\author{
    Ian~M.~Hammond,\authormark{1*}
    Pengning~Chao,\authormark{2}
    Henry~O.~Everitt,\authormark{4,5}
    Rasmus~E.~Christiansen,\authormark{6}
    Alan~Edelman,\authormark{2}
    Francesc~Verdugo,\authormark{3}
    and Steven~G.~Johnson\authormark{2}
}

\address{
\authormark{1}Electrical Engineering and Computer Science Department, Massachusetts Institute of Technology, Cambridge, MA 02139, USA\\
\authormark{2}Department of Mathematics, Massachusetts Institute of Technology, Cambridge, MA 02139, USA\\
\authormark{3}Department of Computer Science, Vrije Universiteit Amsterdam, De Boelelaan 1111, 1081 HV Amsterdam, The Netherlands\\
\authormark{4}U.S. Army Research Laboratory-South, Rice University, Houston, TX 77005, USA\\
\authormark{5}Department of Physics, Duke University, Durham, NC 27708, USA\\
\authormark{6}Department of Civil and Mechanical Engineering, Technical University of Denmark, Nils Koppels Allé, Building 404, 2800 Kongens Lyngby, Denmark\\
}

\email{\authormark{*}ihammond@mit.edu}

\begin{abstract}
Topology optimization (TO) of 3d surface-enhanced Raman scattering (SERS) substrates faces challenges in managing field singularities and modeling orientation-averaged anisotropic molecules. We present 3d TO for manufacturable SERS substrates maximizing spatially averaged signals from randomly oriented, anisotropic molecules, for elastic and inelastic scattering. A new trace formulation provides a closed-form rotational average of anisotropic Raman tensors, which are \emph{not} equivalent to isotropic molecules due to the tensor nonlinearity. Optimized Ag and Si$_3$N$_4$ devices show that lengthscale constraints are sufficient to suppress designs that rely on unphysical mathematical field divergences at sharp corners. Metal designs deliver broadband enhancement and remain robust to typical Raman shifts, whereas dielectric designs yield narrower, $Q$-limited gains that are inferior to the metallic designs for $Q \lesssim 500$. Our approach readily incorporates additional physics, such as a nonlinear damage model. Together, these results provide a practical route to improved manufacturable SERS substrates and extend naturally to other distributed-emitter design problems.
\end{abstract}

\section{Introduction}
\label{sec:introduction}

Raman spectroscopy measures inelastic scattering: a pump drives a polarization radiating at a frequency-shifted emission wavelength; in SERS,  engineered metallic or dielectric nanostructures can strongly amplify both the local pump field and the out-coupling of the emitted field~\cite{jones2019raman,garrell1989surface,langer2019present,long2002raman}. The detected signal is typically an incoherent sum over many spatially and orientationally random emitters, requiring an ensemble-averaged figure of merit. Prior SERS structures range from sphere/dimer arrays to hand-designed metasurfaces~\cite{rycenga2010understanding,hao2004electromagnetic,zhang2012surface}, and recent work has begun to target Raman objectives by inverse design~\cite{christiansen2020inverse,PanCh21,yao2023designing}, revealing the existence of vastly superior SERS geometries. However, significant challenges remain for 3d systems. First, 3d field singularities at sharp corners are stronger than in 2d, challenging standard design formulations~\cite{andersen1978field,idemen2003confluent,meixner1972behavior}---specifically, the integrated signal diverges for localized conical tips sharper than a critical angle (where fields scale as~$r^{t-1}$ with distance $r$ from the tip and geometry--angle exponent $t < 1/4$). Second, we show that the orientational average of realistic anisotropic molecules does \emph{not} reduce to an effective isotropic model, due to a term that arises from the nonlinear averaging of $\alpha$.

We demonstrate topology optimization (\secref{sec:to-numerics}) of manufacturable 3d SERS metasurfaces~\cite{jones2019raman,garrell1989surface,langer2019present}, averaged over both positions and orientations of the Raman molecules, for both metallic and dielectric surfaces, with either 3d or 2d degrees of freedom, including the effect of ``hotspots'' and nonlinear damage. We find that imposing manufacturing minimum feature size constraints (\secref{sec:hotspots}) is sufficient to avoid field singularities from sharp tips (\secref{sec:hotspots}), which would otherwise lead to diverging Raman emission~\cite{andersen1978field,idemen2003confluent,meixner1972behavior,le2006rigorous}; including nonlinear damage effects~\cite{yang2013ultraviolet,fang2008measurement} can further suppress high field intensities (\secref{sec:nonlinear-damage}). Spatial averaging for randomly distributed molecules is handled efficiently by a trace-based technique~\cite{yao2022trace,polimeridis2015fluctuating,rodriguez2012fluctuating} that requires only 1--3 Maxwell solves to obtain the average over all positions (\secref{sec:trace}), previously demonstrated for 2d-Raman~\cite{yao2023designing} and 3d-scintillation~\cite{roques2022framework} inverse design. However, Raman molecules generally have anisotropic polarizability~\cite{long2002raman}, and we show analytically that averaging over orientations is \emph{not} equivalent to an effective isotropic polarizability (Appendix~A); nevertheless, we find that surfaces optimized for isotropic molecules work well for anisotropic molecules and vice versa, with spectral deviations staying within $\sim$20\% when all designs are evaluated using the anisotropic figure of merit (\secref{sec:anisotropy-impact}). We present results (\secref{sec:results}) for both metallic surfaces, which exhibit high field concentrations and broadband plasmonic resonances~\cite{zhang2012surface,rycenga2010understanding}, and lossless dielectric surfaces, which can support very high-$Q$ resonances~\cite{koshelev2020subwavelength,fang2024million} at the expense of bandwidth and robustness (regularized here by artificial damping~\cite{liang2013formulation}). For $Q$ below $\approx 500$, we find that metal surfaces are superior, especially when there is a large split between the pump and emission wavelengths (\secref{sec:inelastic}). Recent theoretical bounds indicate that even lossy materials can, in principle, achieve unbounded Raman response via arbitrarily high-$Q$ delocalized resonances (in the ideal limit, neglecting quantum effects and fabrication perturbations)~\cite{chao2025sum}, but such states are excluded when a minimum bandwidth is imposed and the period is chosen strategically. We also find that fully freeform 3d surfaces are not significantly better than (more manufacturable) 2d patterns for our subwavelength-thickness surfaces (\secref{sec:hotspots}), whereas the traditional SERS geometry of spherical metal particles~\cite{rycenga2010understanding,hao2004electromagnetic} performs significantly worse. Finally, imposing mirror symmetries (\secref{sec:geometryparams}) and optimizing only a single emission polarization (\secref{sec:metal-results}) do not significantly degrade performance. We believe that a similar computational and theoretical framework should benefit many other Raman and incoherent nonlinear-emission design problems, but additional analysis is needed to probe very high-$Q$ regimes (\secref{sec:conclusion}).

We cast the problem as large-scale topology optimization of the permittivity distribution, solved with gradient-based methods and adjoint sensitivities (\secref{sec:to-numerics})~\cite{jensen2011topology,bendsoe2013topology,molesky2018inverse,johnson2012notes}. Although Raman effects are nonlinear in the electromagnetic fields, due to the weak emission one can typically employ linear frequency-domain Maxwell solves at the pump and emission while embedding the nonlinear physics in the objective ~\cite{reif2009fundamentals,long2002raman,le2006rigorous,yao2022trace}. We consider both a freeform 3d design and also a more manufacturable ``2d'' design: a 2d mask that is uniformly extruded into 3d to represent a lithographically etched pattern, with explicit minimum-lengthscale control via Helmholtz filtering and constraints~\cite{zhou2015minimum,hammond2021photonic,arrieta2025hyperparameter}.  We show that additional symmetries can be imposed on both the structure and emission that reduce computational cost without degrading Raman performance (\secref{sec:geometryparams}). The precise mathematical operators and orientation-averaging are described in \secref{sec:trace}, where we obtain a surprising result that averaging an anisotropic molecule over all orientations yields a closed-form expression that is \emph{not} equivalent to any isotropic-molecule response.

    \begin{figure}[tb]
        \centering
        \includegraphics[width=0.6\textwidth]{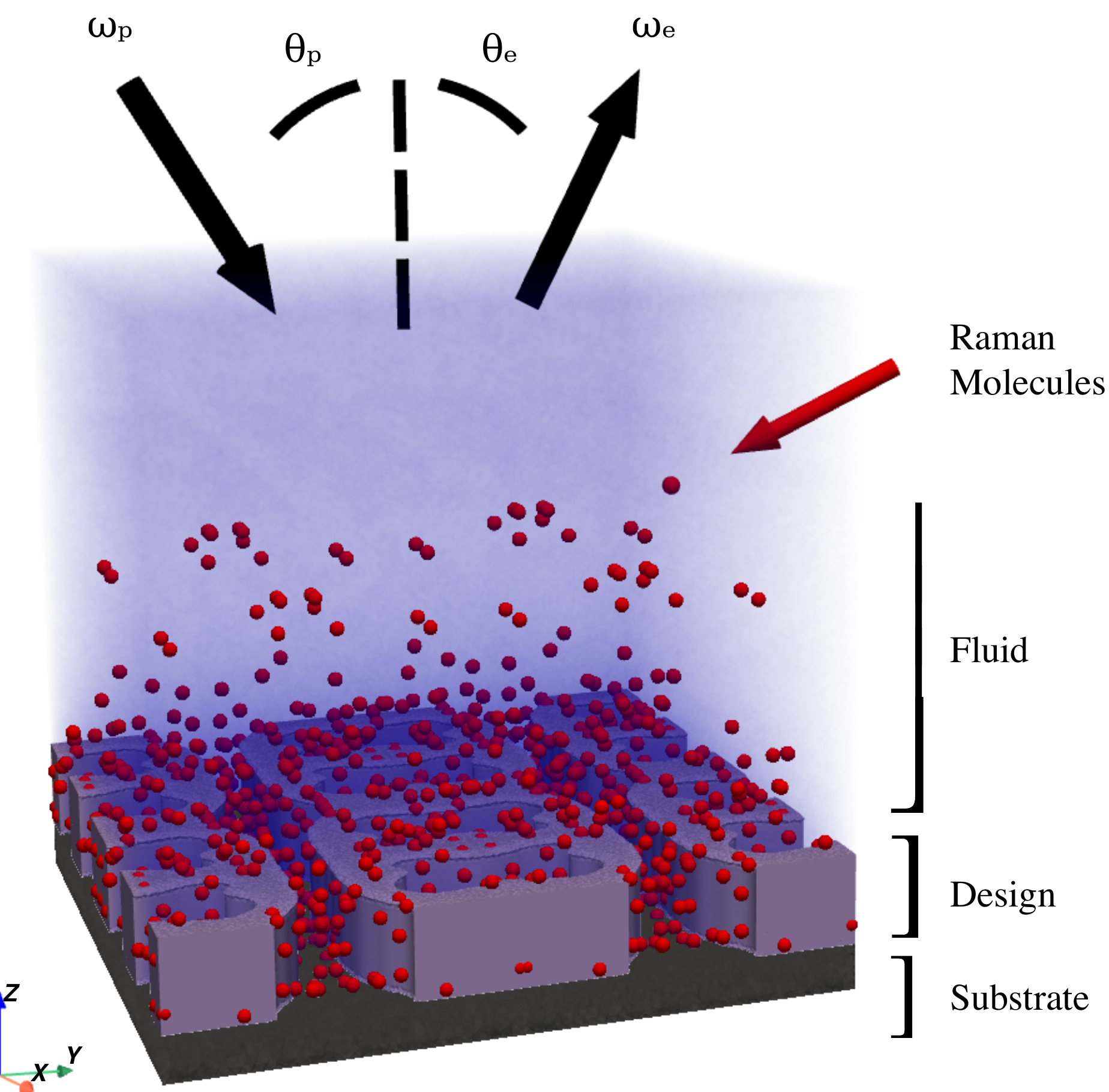} 
        \caption{An overview of the 3d SERS setup illustrates the substrate, design layer, and fluid layer containing Raman-active molecules. An incident pump wave (frequency $\omega_\text{p}$, angle $\theta_\text{p}$) excites the molecules, leading to emitted Raman light (frequency $\omega_\text{e}$, angle $\theta_\text{e}$).}
        \label{fig:figure-1-overview}
    \end{figure}

\section{Methods}
\label{sec:methods}
    \subsection{Raman Scattering Model and Trace Formulation}
        \label{sec:trace}
        The interaction of light with Raman-active molecules, as illustrated in \figref{fig:figure-1-overview}, is modeled using Maxwell's equations for the incident pump electric field ($\vect{E}_\text{p}$) at frequency $\omega_\text{p}$ and the Raman-scattered emitted field ($\vect{E}_\text{e}$)  at frequency $\omega_\text{e}$. However, for emission in a particular direction and polarization, one can equivalently compute the emission by solving for a reciprocal field $\vect{E}'_\text{e}$ resulting from an \emph{incident} field from the emission direction/polarization~\cite{yao2023designing}. 
        The SERS objective function, $g$, is the ensemble average of the Poynting flux, $P_{\vect{r}_0, \alpha}$, radiated by molecules at positions $\vect{r}_0$ with Raman polarizability tensor $\alpha$, averaged over all molecular positions and orientations $Q \in \text{SO}(3)$ (3d rotation group) ~\cite{long2002raman}, which we assume are uniformly distributed (e.g.~by diffusion in a fluid) in some region of space. 
        Accounting for non-uniform molecular distributions only requires augmenting the prescribed distribution in the objective operator; in particular, molecular aggregation~\cite{pei2025enrichment,Wu2026LightField} can be modeled as a spatially varying concentration weight without changing the overall optimization workflow~\cite{yao2022trace}.
        This is computed exactly using an overlap integral between $\vect{E}_\text{p}$ and $\vect{E}'_\text{e}$~\cite{yao2023designing}, which is vastly more efficient than averaging individual emitters.
        For isotropic molecules with polarizability $\alpha_0$, elastic scattering ($\omega_\text{e} \approx \omega_\text{p}$),  and emission in the same direction and polarization as the pump, one has $\vect{E}'_\text{e} \approx \vect{E}_\text{p}$ and the objective simplifies to $\int_{\Omega}|\alpha_0|^2 |\vect{E}_\text{p}|^4 d\Omega$ where $\Omega$ is the region in which the molecules are uniformly distributed~\cite{yao2023designing, le2006rigorous, PanCh21}. More generally, for isotropic molecules with inelastic scattering ($\omega_\text{e} \ne \omega_\text{p}$) and/or emission into a different direction or polarization than the pump, the $|\vect{E}_\text{p}|^4$ integrand is replaced by $|\vect{E}_\text{p}(\vect{x})^* \cdot \vect{E}'_\text{e}(\vect{x})|^2$~\cite{yao2023designing,hammond2024designing}. Even more generally, the Raman polarizability $\alpha$, yielding a dipole source $\vect{p} = \alpha \vect{E}_\text{p}$ at $\omega_\text{e}$,  is an anisotropic $3 \times 3$ tensor~\cite{long2002raman}. We show in Appendix~\ref{sec:appendix-anisotropy}, below, that averaging the emitted power over all molecule orientations (all rotations of $\alpha$) is nontrivial due to the nonlinearity in $\alpha$, yielding an integrand that is different from that of an isotropic molecule, but remains computationally simple.

        We employ a finite-element method (FEM) model for discretizing Maxwell's equations~\cite{jin2015finite} into a matrix equation  $\mat{M} \vect{u} = \vect{b}$ relating discretized currents $\vect{b}$ and the resulting electric fields $\vect{u}$ by a sparse matrix $\mat{M}$.  One such equation $\mat{M}_\text{p} \vect{u}_\text{p} = \vect{b}_\text{p}$  at $\omega_\text{p}$ is solved to obtain the pump field $\vect{u}_\text{p}$. Similarly, at $\omega_e$ we solve a reciprocal problem for the emission (source term $\vect{b}_e=\mat{A} \vect{u}_\text{p}$ from the detector mode with $\mat{A}$ a sparse matrix representing the action of the Raman polarizability tensor as described below) to obtain $\vect{u}_e$ (or $\vect{v}_k$ in Eq. 2) via the reciprocity principle. The spatially and orientationally averaged objective $g$ can then be computed using the trace identity~\cite{yao2022trace, lax2007linear}:
        \begin{equation}\label{eq:trace_objective}
            g = \text{Tr} [ \mat{M}_\text{e}^{-\dagger} \mat{O} \mat{M}_\text{e}^{-1} \mat{B} ] \, ,
        \end{equation}
        Here, $\mat{M}_\text{e}$ is the sparse Maxwell matrix at $\omega_\text{e}$, $\mat{O}$ is the output projection operator representing the detection channel(s), and $\mat{B} = \langle \mat{A} \vect{u}_\text{p} \vect{u}_\text{p}^\dagger \mat{A}^\dagger \rangle_{\alpha, \vect{r}_0}$ is the ensemble-averaged correlation matrix of the emission sources. As reviewed below, this trace can be evaluated efficiently for emission into a single direction because $\mat{O}$ is low rank~\cite{yao2022trace,yao2023designing}.

        A key contribution of this work is the formulation of the correlation matrix $\mat{B}$ for anisotropic emitters. For a complex-symmetric (reciprocal) Raman polarizability tensor $\alpha$ and a complex vector $\vect{x}$ (representing the components of $\vect{E}_\text{p}$ at a molecular site), the rotationally averaged dyadic term, critical for constructing $\mat{B}$, is found to be (see Appendix A for derivation):
        $\langle Q \alpha Q^\top \vect{x} \vect{x}^\dagger Q \alpha^\dagger Q^\top \rangle_Q = (\alpha_\|^2 - \alpha_\perp^2) \vect{x} \vect{x}^\dagger + \alpha_\perp^2 ( \vect{x}^\dagger\vect{x} \mat{I} - 2i \text{Im}[\vect{x}\vect{x}^\dagger] )$.
        The terms $\alpha_\|^2$ and $\alpha_\perp^2$ represent effective parallel and perpendicular polarizability components arising from the $\text{SO}(3)$ (the group of 3d rotations) averaging, expressed in terms of the rotation-invariant quantities $|\Tr(\alpha)|^2$ and $\Tr(\alpha\alpha^\dagger)$. The Hermitian semidefinite matrix on the right-hand side is generally not equivalent to the $\alpha_0^2 \vect{x}\vect{x}^\dagger$ matrix that would arise for any isotropic polarizability---the second $\alpha_\perp^2(\cdots)$ term captures crucial phase and polarization information necessary for accurately modeling a rotation-averaged anisotropic response, and the difference arises from fact that we are averaging a quantity that is nonlinear in $\alpha$.

        The output operator $\mat{O}$ depends on the detection scheme, but a key enabling factor for computational efficiency is that it is low rank~\cite{yao2022trace}.  In particular, for computing the power emitted into a single direction with two possible polarizations, we have a rank-2 matrix  $\mat{O} = \vect{o}_x \vect{o}_x^\dagger + \vect{o}_y \vect{o}_y^\dagger$  consisting of  projections onto the $x$- and $y$- polarized output modes $\vect{o}_x$ and $\vect{o}_y$ (derived from mode orthogonality~\cite{yao2022trace,snyder1983optical}).  However, we find in \secref{sec:metal-results} that the emission is typically dominated by the same polarization ($y$) as the pump, so we can approximate the response by a rank-1 matrix  $\mat{O} = \vect{o}_y \vect{o}_y^\dagger$.   In general, for any rank-$k$ matrix  $\mat{O} = \sum_j\vect{o}_j \vect{o}_j^\dagger$, the trace formula \eqnref{eq:trace_objective} for $g$ simplifies to~\cite{yao2022trace}
        \begin{equation}\label{eq:trace_objective_lowrank}
            g = \sum_{j=1}^k \vect{v}_j^\dagger \mat{B} \vect{v}_j \, ,
        \end{equation}
where $\vect{v}_k = \mat{M}_\text{e} ^{-\dagger} \vect{o}_k = \overline{\mat{M}_\text{e} ^{-T} \overline{\vect{o}_k}} = \overline{\mat{M}_\text{e} ^{-1} \overline{\vect{o}_k}}$  is the ``reciprocal'' solution $\vect{E}'_\text{e}$ treating the output mode $\vect{o}_k$ as a source term, exploiting reciprocity $\mat{M}_\text{e} ^{T} = \mat{M}_\text{e}$~\cite{yao2022trace}.  This corresponds to two Maxwell solves in our 2-polarization case, and only one Maxwell solve in the 1-polarization approximation, in addition to the Maxwell solve for the pump. In the elastic 1-polarization case with $\mat{M}_\text{e} = \mat{M}_\text{p}$, the reciprocal solution is equivalent to the pump solution and we only require a single Maxwell solve.
        Gradients of the objective function with respect to the design variables are computed efficiently using the adjoint method, a standard technique in large-scale optimization~\cite{johnson2012notes, molesky2018inverse}.  For every Maxwell solve to compute $g$, we require one additional ``adjoint'' Maxwell solve to obtain the gradient $\nabla g$.

    \subsection{SERS Substrate Topology Optimization: Numerical Implementation}
        \label{sec:to-numerics}
        The design variables in our TO framework are an artificial piecewise-constant ``density'' field $\rho(\vect{r})\in[0,1]$, defined on Cartesian grid, from which we obtain the material permittivity distribution $\epsilon(\vect{r})$ by existing filter-project method ~\cite{jensen2011topology, bendsoe2013topology, hammond2025unifying}.  In particular, given a scalar density field $\rho(\vect{r})\in[0,1]$ , we low-pass filter to obtain a smoothed field $\tilde{\rho}(\vect{r})$ (that regularizes the problem by introducing a minimum lengthscale $\sim 20$\,nm).  Additional lengthscale/manufacturing constraints are discussed in Sec.~\ref{sec:hotspots}. This is then projected to a binarized density $\hat{\rho}$  with a steepness parameter $\beta$, using a recent differentiable subpixel-smoothed projection (SSP) scheme $\hat\rho(\tilde \rho, \Vert \tilde \rho\Vert)$ that allows us to increase $\beta$ to $\infty$ (guaranteeing an almost-everywhere binary structure) without losing differentiability~\cite{hammond2025unifying}; all optimizations increased $\beta$ in four epochs $\beta \in  \{ 8.0, 16.0, 32.0, \infty \}$.  The projected density $\hat{\rho}\in[0,1]$ is converted into an interpolated permittivity $\varepsilon$ at each point, using a specialized interpolation scheme developed for metals to avoid artificial bulk-plasmon resonances at intermediate $\hat{\rho}$~\cite{christiansen2019non}.  
        We considered both freeform 3d structures and lithography-like 2d-patterned layers, i.e. $\rho(x,y,z)$ or $\rho(x,y)$. In the former 3d case, $\rho(x,y,z)$ is defined on the same finite-element mesh as the electromagnetic fields, and is low-pass filtered using the modified Helmholtz method~\cite{lazarov2011filters}.  In the latter 2d case, $\rho(x,y)$ is defined on a 2d Cartesian grid, with $\tilde\rho$ computed by a conic filter~\cite{hammond2021photonic}, and then bilinearly interpolated into the quadrature points of the 3d finite-element mesh to obtain $\varepsilon(x,y,z)$ with piecewise constant cross section. (Gradients are backpropagated to the 2d grid through this interpolation process~\cite{hammond2024thesis_ian,hammond2025unifying}.) 
        Optimization is performed using the algorithm of Conservative Convex Separable Approximations (CCSA), a gradient-based algorithm well-suited for problems with many variables and inequality constraints~\cite{svanberg2002class}, implemented via the NLopt library (with the \texttt{CCSAQ} option)~\cite{NLopt}.
        For dielectric optimizations, an artificial material loss $\kappa = 1/(2Q)$ is introduced to broaden resonances in early epochs (to make them easier to find) and later bound the resonant quality factors ($Q$)~\cite{liang2013formulation}. The value of $\kappa$ per epoch is successively decreased as $\{ 0.1, 0.01, 0.001\}$, coinciding with $\beta$ increases, with the final value typically yielding $Q\approx$~100--200 (out of a maximum $\kappa$-limited $Q_{\text{max}} \approx 500$).  We set $\kappa=0$ when evaluating the final optimized design.
        Initial conditions for the design variables $\rho$ are typically random (uniform random values $\in[0,1]$ followed by low-pass filtering with the same $20$\,nm lengthscale to obtain a more regular structure) to break any unintended symmetries (unless specific symmetries are desired) and to avoid trivial high-symmetry solutions (e.g.~$\rho = 0$ or $\rho =1$ everywhere).

        \subsubsection{Physical and Geometrical Parameters}
        \label{sec:geometryparams}
        The materials considered are silver (Ag, optical constants from~\cite{johnson1972optical_Ag}) on a silver substrate, and silicon nitride (Si$_3$N$_4$, constants from~\cite{luke2015broadband_SiN}) on a silicon dioxide (SiO$_2$) substrate (constants from~\cite{malitson1965interspecimen_SiO2}). The surrounding medium is water (n $\approx$ 1.33~\cite{hale1973optical_H2O}), corresponding to Raman detection for molecules suspended in a thin fluid layer.  The Raman molecules were assumed to be uniformly randomly distributed within the fluid layer, extending a thickness 200\,nm above the design region, with the fluid and molecules also penetrating into any holes in the design region.
        Optimized designs are heavily influenced by the specific material choice as explored in our previous works~\cite{hammond2024thesis_ian, christiansen2020inverse}.
        The thickness of the design region is  100\,nm for metals and 200\,nm for dielectrics.  The period of the SERS substrate is chosen in the metallic case to target plasmonic resonances (e.g., surface plasmon polaritons~\cite{zhang2012surface}), following a method similar to Yao et al.~\cite{yao2023designing} based on the material properties at the operating wavelength; the precise values are given in the results sections below.
        Meshing employed tetrahedral zeroth-order Nedelec elements in the FEM solver Gridap.jl~\cite{Badia2020,Verdugo2022}, with a 6th-order quadrature rule, where element sizes near the design region were approximately 2\,nm for metals (to accurately resolve the skin depth~\cite{hassan2021study}) and 5\,nm for dielectrics. Away from the active design region, a coarser 10\,nm mesh is used to reduce computational load.  Simulations were performed on the MIT SuperCloud high-performance computing cluster using the Pardiso sparse-direct solver~\cite{schenk2004solving}. Typical metal designs required approximately three days to complete 800 iterations using 40 CPUs and 186~GB RAM.
        
        The computational cell size was further reduced by a factor of~4 by imposing two mirror-symmetry planes in $x$ and $y$. To assess the impact of this choice, optimizations were also performed without imposing mirror symmetries. Optimized structures showed negligible change when symmetry constraints were removed, supporting the use of symmetry to reduce computational cost without compromising performance. Even without imposing symmetry, we observed that at least one mirror symmetry tends to arise in the optimized structure, even if the initial structure is non-symmetric.

    \subsection{Hotspots and lengthscale constraints}
        \label{sec:hotspots}
        SERS topology optimization requires a minimum-lengthscale constraint, both by the density smoothing $\tilde \rho$  above and by more explicit manufacturing constraints~\cite{zhou2015minimum,arrieta2025hyperparameter,hammond2021photonic}, to suppress field singularities (hotspots) arising from sharp tips.  In particular, the $\int|\vect{E}|^4$ objective otherwise favors geometric singularities where fields diverge (with increasing mesh resolution) in a non-integrable way, leading to an unattainable diverging Raman response~\cite{yao2023designing}.  
        In 3d, this effect is more pronounced than in previous 2d studies~\cite{yao2023designing}: the field near a 3d conical tip scales as $r^{t-1}$ for the radius $r$ and tip parameter $t(\varepsilon,\phi)$, leading to an integral $\int |\vect{E}|^4 r^2 dr \sim r^{4t-1}$, which diverges for $t < 1/4$~\cite{andersen1978field,idemen2003confluent,meixner1972behavior,yao2023designing}. In contrast, a 2d corner has weaker $r^{4t-2}$ scaling, and only leads to a diverging $\int|\vect{E}|^4$  for metals at a particular corner angle~\cite{yao2023designing}.  
        Without any regularization, we find that TO leads to many artificial hotspots~\cite{hammond2024thesis_ian}, even from the ``roughness'' of the FEM mesh itself. Arbitrarily sharp tips are not readily manufacturable, and ultimately the field divergences are regularized by nonlocal quantum effects at lengthscales $\lesssim 10$\,nm~\cite{eguiluz1976hydrodynamic, boardman1982electromagnetic}. These do not converge with resolution. 
        We employ a common minimum-lengthscale constraint formulation~\cite{zhou2015minimum}, recently updated to simplify its implementation in conjunction with SSP~\cite{hammond2025unifying,arrieta2025hyperparameter}, imposing a minimum linewidth of $20$\,nm on both solid and void regions.
        Fortunately, we find that by imposing a lengthscale regularization, and especially by imposing explicit manufacturing constraints on the minimum lengthscale (which can otherwise be violated by the nonlinear projection $\hat{\rho}$ of the smoothed density $\tilde \rho$~\cite{hammond2025unifying}), we can force TO to restrict itself to physically attainable SERS structures lacking sharp tips.   (Nor does the optimization seem to get ``stuck'' at intermediate steps violating the constraints.)   For 2d-patterned constant cross-section 3d structures, there are $90^\circ$ sharp ``2d'' corners at the top and bottom edges of the pattern, but this does not lead to a diverging  $\int|\vect{E}|^4$   for either our dielectric materials (where 2d singularities are always integrable~\cite{yao2023designing}) or our metallic material (where this is the wrong angle for a non-integrable singularity~\cite{yao2023designing}).
        Our previous work~\cite{hammond2024thesis_ian} explores experiments varying minimum-lengthscale constraints; as the prescribed length scale decreases, optimized metallic features sharpen. 

\section{Results}
\label{sec:results}

    \subsection{Nanoplasmonic Results (Metallic Substrates)}
        \label{sec:metal-results}
\begin{figure}[tb]
            \centering
            \includegraphics[width=1.0\textwidth]{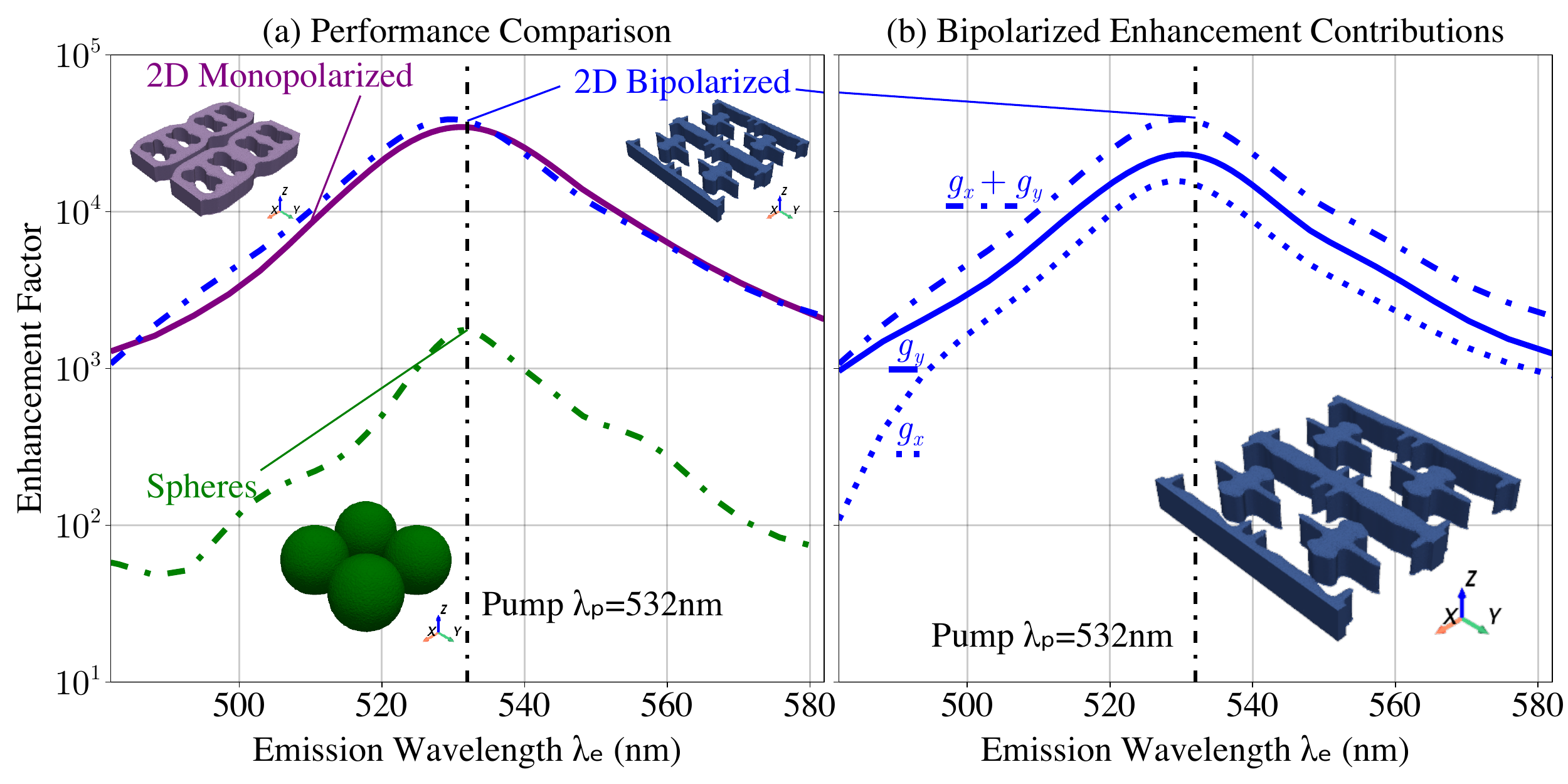} 
            \caption{Performance and polarization for 2d metallic SERS TO. (a) Enhancement factor for designs optimized for monopolarized (y-output) and bipolarized (x- and y-output) emission, compared to optimized sphere arrays with 20\,nm gaps. (b) Contributions to enhancement ($g_x, g_y$) for the bipolarized 3d design, illustrating the dominance of one polarization.}
            \label{fig:figure-2-results}
        \end{figure}

        The performance of 3d optimized metallic SERS substrates, designed using the 2d-DOF formulation, is presented in \figref{fig:figure-2-results}. The  period used for the optimized spectra was 184\,nm, selected to target the primary plasmonic resonance at the pump wavelength. The pump and emission directions are both normal to the surface.  We find that optimization yields more than an order of magnitude improvement over an array of optimized-period spheres, and that the emitted power is almost entirely in the same polarization as the pump light. All subsequent spectral plots show the emission wavelength being swept unless otherwise noted. All optimizations assume normal incidence of the pump wave.
        
        \figref{fig:figure-2-results}(a) compares the SERS enhancement factor for designs optimized for monopolarized ($y$-polarized output) and bipolarized ($x$- and $y$-polarized output) emission, for a $y$-polarized pump, against benchmark sphere arrays. These arrays consist of radius-82\,nm spheres with a 20\,nm gap (period 184\,nm), matching the TO lengthscale. The TO designs consistently and significantly outperform these simple sphere arrays, highlighting the power of freeform 2d-pattern optimization. \figref{fig:figure-2-results}(b) dissects the bipolarized (dual-channel) enhancement, showing that the $y$-polarized contribution ($g_y$) typically dominates the $x$-polarized one ($g_x$). Furthermore, our ``monopolarized'' TO designs optimized for single-channel $y$-output achieve a total enhancement $g = g_x + g_y$ comparable to the bipolarized designs, suggesting that optimizing for a single, strategically chosen polarization can be an effective approach (and reduces the computational cost by eliminating the Maxwell solves for the $x$ polarization), and leads to a qualitatively simpler-looking structure.
        
The issue of local optima is pertinent in TO. For our 2d-DOF metal designs (single-channel), optimizations initiated from 10 different random starting conditions yielded a standard deviation in the final objective function value of approximately 55\%. Despite this variance in absolute performance, a striking convergence in morphology was observed: most designs evolved to share the ``rounded corners in $y$-direction'' structural motif, indicating a robust underlying design principle favored by the optimization for this specific problem.

Resonant peak locations are inherently sensitive to structural details and surface plasmon effects, but the order-of-magnitude response is robust.  In addition to manufacturing uncertainty, there is also discretization error introduced by the finite mesh resolution and the 1-voxel layer of ``gray'' (intermediate/interpolated) materials introduced at interfaces by the SSP projection.  To characterize this effect, we extracted the level-set contour of the optimized surfaces and generated an interface-conforming mesh to recompute the Raman enhancement~\cite{sigmund2013topology}.  We found that the resonant wavelength shifted by $\approx 5$\,nm and that the peak enhancement actually \emph{increased} by several times, reflecting the sensitivity of surface plasmons to surface details.  However this is compensated somewhat by the fact that realistic silver surfaces will typically be coated  by a thin ($\sim 2$\,nm) layer of oxide; in high-resolution 2d calculations, we find that such an oxide layer typically reduces enhancement by a factor of $\sim 2$.  Ultimately, however, Raman enhancement is so large that it is dominated by order-of-magnitude concerns.

    \subsection{Dielectric Results}
        \label{sec:dielectric-results}
\begin{figure}[tb]
            \centering
            \includegraphics[width=1.0\textwidth]{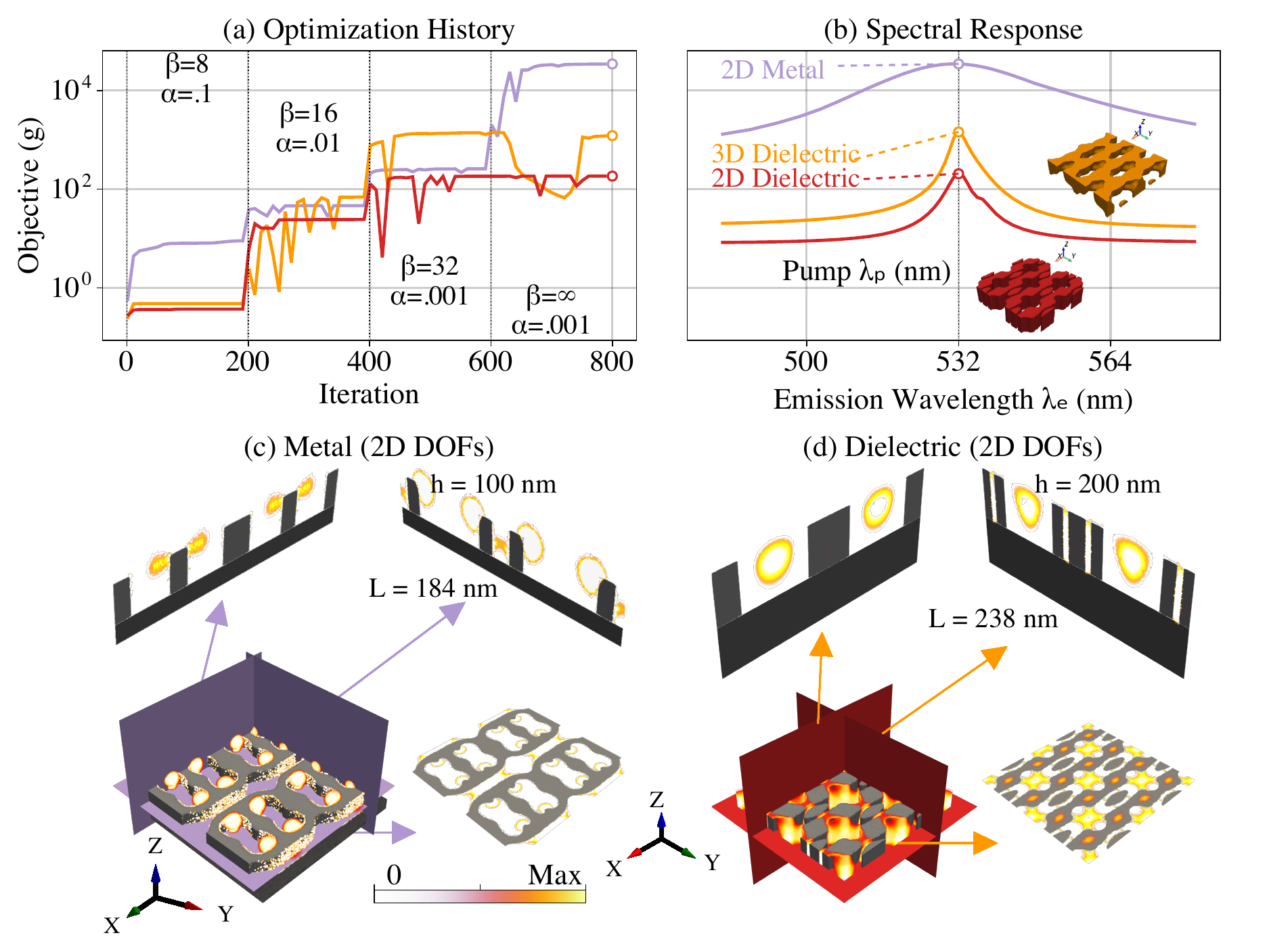} 
            \caption{3d dielectric SERS TO with 2d-DOFs. (a) Objective $g$ vs. iteration, showing $\beta$ and $\kappa$ epochs for the 2d metal (blue), 3d dielectric (orange), and 2d dielectric (red) optimizations in this work. (b) Spectral response $g$ vs. wavelength for the optimized 3d dielectric design compared to the 2d dielectric design. Insets: (b, top right) 3d and 2d optimized dielectric geometries compared against 2d metallic; (c) $|\vect{E}|^4$ (yellow) and material (gray) cross sections for the 2d optimized metal, highlighting the concentration near the surface of the device; (d) $|\vect{E}|^4$ cross-section for the 3d optimized dielectric, illustrating field concentration in air.}
            \label{fig:figure-3dielectric}
        \end{figure}

        The optimization of 3d dielectric substrates (Si$_3$N$_4$ on SiO$_2$) using the 2d-DOF formulation is summarized in \figref{fig:figure-3dielectric}.  Figure \ref{fig:figure-3dielectric} compares an optimized 3d-freeform dielectric structure, a 2d-patterned dielectric, and the optimized 2d metal surface from the previous section, all optimized for ``monopolarized'' operation (maximizing $y$-polarized emission for a $y$-polarized pump).
        The dielectric designs were simulated with an in-plane period of 238\,nm, chosen to place a guided-mode resonance near the operating wavelength while satisfying the same fabrication constraints as the metal designs.  As discussed below, these dielectric surfaces were optimized for a moderate $Q$ of 200--300 ($\approx 0.5$\% bandwidth), and we find that in this regime the metallic surface outperforms the dielectric structures by at least an order of magnitude.  (In principle, an arbitrarily large Raman enhancement can be obtained by arbitrarily high-$Q$ resonances in ideal lossless dielectrics with no manufacturing disorder, e.g.~using bound-in-continuum resonances~\cite{hsu2016bound}, but going to higher $Q$ also trades off robustness and bandwidth.)
        
        \figref{fig:figure-3dielectric}(a) plots the optimization history ($g$ vs.\ iteration), highlighting the $\beta$-schedule and the artificial loss $\kappa$ schedule applied to the 3d dielectric (orange), the 2d dielectric (red), and 2d metal (purple).
        Fabrication-lengthscale constraints are turned on in the final epoch (with $\beta=\infty$).
        The spectral response ($g$ vs.\ wavelength) of the optimized designs are compared in \figref{fig:figure-3dielectric}(b). The 3d structure delivers nearly an order of magnitude higher enhancement than its 2d-patterned counterpart, underscoring the benefit of volumetric degrees of freedom, though both underperform the optimized 2d metal surface.
        
        The geometric insets of \figref{fig:figure-3dielectric}(b) show the 3d and 2d dielectric designs, and the $|\vect{E}|^4$ cross-sections in panels (c--d) illustrate the 2d metal and dielectric designs.  Panels (c--d) also contrast the field localization paths relative to the metal baseline: plasmonic devices concentrate energy near metal surfaces and gap regions, while these dielectric designs concentrate fields within voids (and in general are expected to be less spatially localized than plasmonic resonances).
        
        The desired quality factor ($Q$) of the dielectric resonances, or equivalently the bandwidth of the enhancement, is a critical parameter for SERS enhancement.  As discussed in Sec.~\ref{sec:to-numerics}, we imposed an artificial absorption loss that limited the total $Q$ to be $\le 500$, and then removed this loss to evaluate the final performance in \figref{fig:figure-3dielectric}.  The optimized dielectric surfaces both achieve $Q \approx$~200--300.  In contrast, the metallic resonance (which is limited by the physical absorption) has $Q < 50$, or roughly $10\times$ the bandwidth.  

    \subsection{Impact of Raman-Molecule Anisotropy}
        \label{sec:anisotropy-impact}
        In this section, we investigate the impact of an anisotropic Raman polarizability averaged over all orientations, which necessitates the use of the modified objective from Sec.~\ref{sec:trace} (derived in Appendix~A). In particular, we investigated the common case of a uniaxial Raman polarizability (which only responds to pump fields in one direction)~\cite{long2002raman,Nakamoto2009}.  Figure \ref{fig:figure-4-anisotropy} shows that anisotropy has minimal effect on design and performance, as bounded theoretically below.  This means that, in practice, one can simply optimize a Raman surface for isotropic molecules and it should yield similar enhancement for arbitrary anisotropic polarizabilities (which may not be precisely known \textit{a~priori}).

        \figref{fig:figure-4-anisotropy}(a) illustrates the Raman enhancement versus wavelength for both metallic and dielectric optimized 2d patterns.  In particular, we compared the isotropic designs from the previous sections (optimized for isotropic molecules) to an anisotropic design (optimized for uniaxial molecules, whose polarizability $\alpha$ has a single nonzero diagonal entry), with both designs finally \emph{evaluated} for the anisotropic objective.  The anisotropic optimization is initialized with the isotropic design (after the second $\beta$ epoch) to ensure that it explores a similar local optimum. We also performed experiments without isotropic initialization (all-anisotropic), which yielded qualitatively similar geometries and objectives within 50\%. This comparison quantifies the performance penalty of optimizing for the ``wrong'' isotropic physics if the true Raman molecule is uniaxial.  We find that the anisotropic designs (dashed lines) are within 20\% of the isotropic designs (solid lines), and within 10\% at the peak, well within the numerical and manufacturing uncertainties, for both metallic and dielectric cases, and the structures (at right) are nearly indistinguishable by eye.

        In fact, since the Raman molecules do not affect the electric fields, and the final objective is a simple quadratic function of the entries of the Raman polarizability tensor $\alpha$ (weighted by the fields), it is possible to obtain strict theoretical bounds on how much the enhancement of a given structure can differ between isotropic and anisotropic molecules. The details of this analysis are included in  Appendix~B. We find that, for uniaxial molecules, the enhancement (for any given structure) can differ from the isotropic enhancement by at most 50\%.   We also show that difference of at most $4\times$ can arise for the more exotic case of a traceless polarizability $\Tr \alpha = 0$. In Raman-spectroscopy language, this traceless case is an anisotropic Raman tensor with vanishing ``isotropic invariant'' (mean polarizability) $\tfrac{1}{3}\Tr \alpha = 0$;  these tensors arise from non-isotropic vibrations such as the $E$ and $T_2$ vibrational modes of the tetrahedral molecule $\mathrm{CCl}_4$~\cite{long2002raman,Nakamoto2009,Subr2018}. 
        In the same Raman-spectroscopy language, the strength of the anisotropic contribution is quantified by the ``anisotropic invariant,'' which in our notation is $\frac{1}{2}\left(3\,\Tr\!\left(\alpha \alpha^{\dagger}\right)-\left|\Tr(\alpha)\right|^2\right)$; this quantity vanishes if and only if $\alpha$ is isotropic (proportional to $I$), and it is exactly this coefficient that controls the extra phase-sensitive coupling term in our orientation-averaged response.

    \subsection{Impact of Inelastic Scattering (Frequency Shifts)}
        \label{sec:inelastic}
\begin{figure}[tb]
            \centering
            \includegraphics[width=1.0\textwidth]{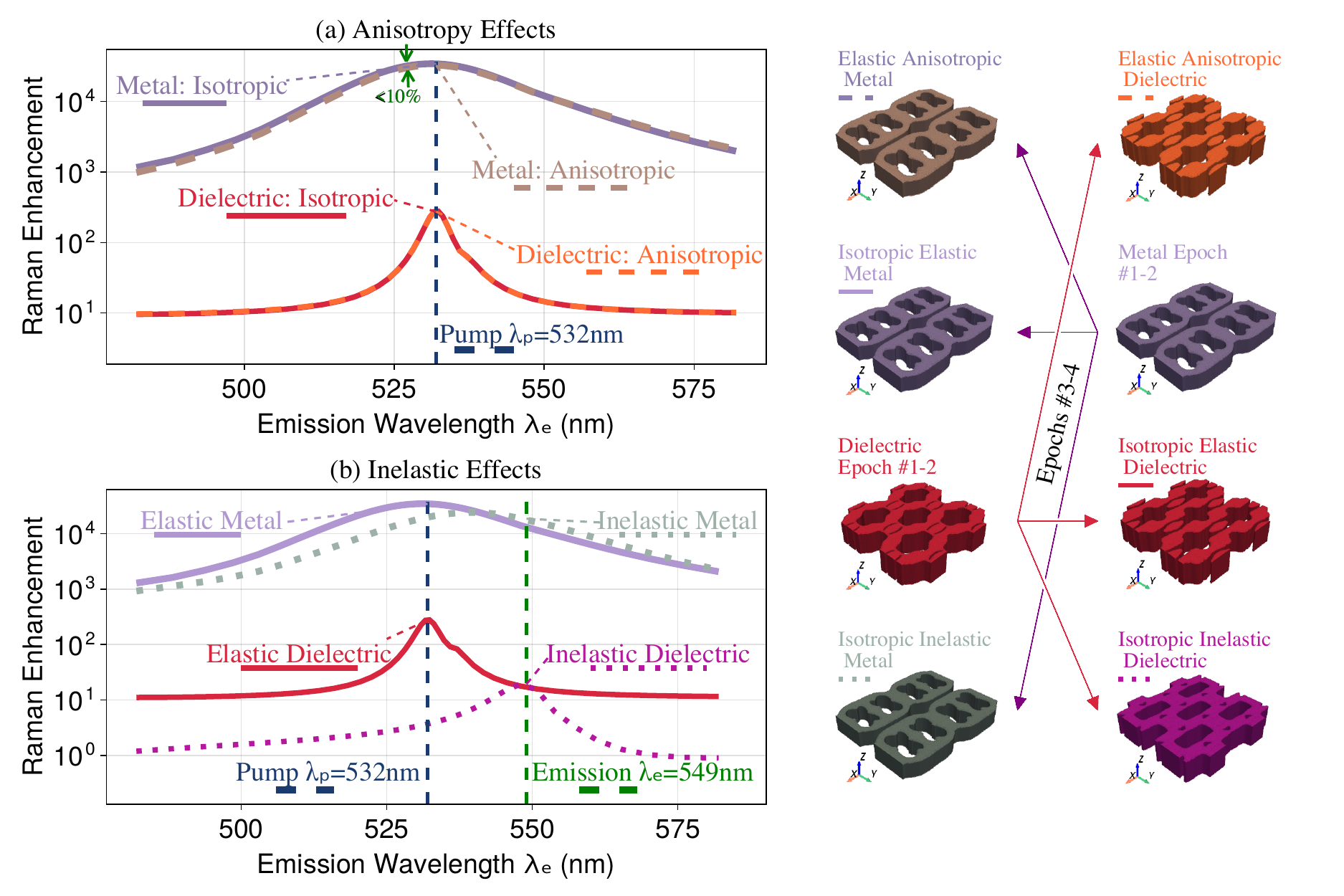} 
            \caption{Impact of (a) material/molecular anisotropy and (b) inelastic Raman scattering on 3d optimized SERS structures (2d-DOFs). All plots show emission wavelength swept, and panel (a) evaluates every spectrum with the anisotropic figure of merit to mimic real molecules while contrasting different design assumptions, revealing that isotropically optimized devices remain within $\sim$20\% of their anisotropic counterparts across the spectrum.  Panel (a) compares Isotropic TO for metal (purple) and dielectric (red) and Anisotropic TO (brown/orange for metal/dielectric). Panel (b) compares Elastic TO (purple/red for metal/dielectric) and Inelastic TO (green/pink for metal/dielectric). All configurations begin elastic and isotropic for the first two epochs to generate comparable local optima. The geometries for all cases are displayed on the right; arrows indicate the final result after the last two epochs under different configurations.}
            \label{fig:figure-4-anisotropy}
        \end{figure}

        This section considers the practical scenario where the pump wavelength $\lambda_\text{p}$ (e.g.~532\,nm) differs significantly from the emission wavelength $\lambda_\text{e}$ (e.g.~549\,nm, corresponding to a Raman shift of $\approx 582$ cm$^{-1}$), focusing on a single emission channel.
        \figref{fig:figure-4-anisotropy}(b) plots the Raman enhancement versus emission wavelength for 2d optimized designs, comparing designs optimized assuming elastic scattering ($\lambda_\text{p} = \lambda_\text{e}$) with those optimized  for specific inelastic  cases. 
        As is done with the anisotropic comparison above, the inelastic optimization is initialized with the isotropic design to ensure that it explores a comparable local optimum. 
        
        For metallic substrates, designs optimized under the elastic-scattering assumption exhibit only minor degradation when evaluated for inelastic scattering (i.e., when $\lambda_\text{e} \neq \lambda_\text{p}$). This robustness can be attributed to the broadband (low-$Q$) plasmonic response of metallic nanostructures~\cite{zhang2012surface}. Structures optimized specifically for inelastic scattering are morphologically very similar to those optimized elastically, with only a slight shift in the center of the resonant spectral response, as seen in the inset geometries.

        In contrast, dielectric designs optimized for elastic scattering suffer a significant performance reduction when evaluated for inelastic scattering, reflecting their typically narrow (high-$Q$) resonant bandwidths. As shown in \figref{fig:figure-4-anisotropy}(b), designs optimized specifically for inelastic conditions retune the main resonance to accommodate the emission wavelength, but their peak enhancement remains comparable to the elastic design evaluated at the pump wavelength. Further attempts (not shown) to simultaneously support strong resonances at both pump and emission wavelengths produced qualitatively different, more complex geometries without delivering superior overall SERS performance. Better doubly resonant dielectric designs may be possible by optimizing a larger unit cell (Sec.~\ref{sec:highq}). These observations suggest that dielectric SERS substrates are less suitable for applications involving large Raman shifts unless  sophisticated high-$Q$ multi-resonant designs can be achieved (also requiring precise fabrication), whereas metal-based designs can often treat common Raman shifts as a negligible perturbation due to their broader spectral features.

    \subsection{Nonlinear-Damage Regularization}
        \label{sec:nonlinear-damage}
        While the primary approach in this work relies on material filtering (with a filter radius $R_\mathrm{f}=20$\,nm) and length-scale constraints inherent in the 2d-DOF formulation to manage geometric singularities, we also explored an additional physical regularization. This exploration is presented not as the central mechanism for hotspot control in our main findings—as the aforementioned geometric constraints were found to be largely sufficient for preventing unphysical mathematical singularities~\cite{hammond2024thesis_ian}—but rather as a demonstration of the framework's adaptability to incorporate other real-world physical effects. (Other regularization strategies, such as direct field-intensity smoothing or nonlocal material models, were also considered in our previous exploratory work~\cite{hammond2024thesis_ian},  including ad hoc exclusion of fields close to the surface for the FOM \cite{vester2017topology}.)
        
\begin{figure}[tb]
            \centering
            \includegraphics[width=1.0\textwidth]{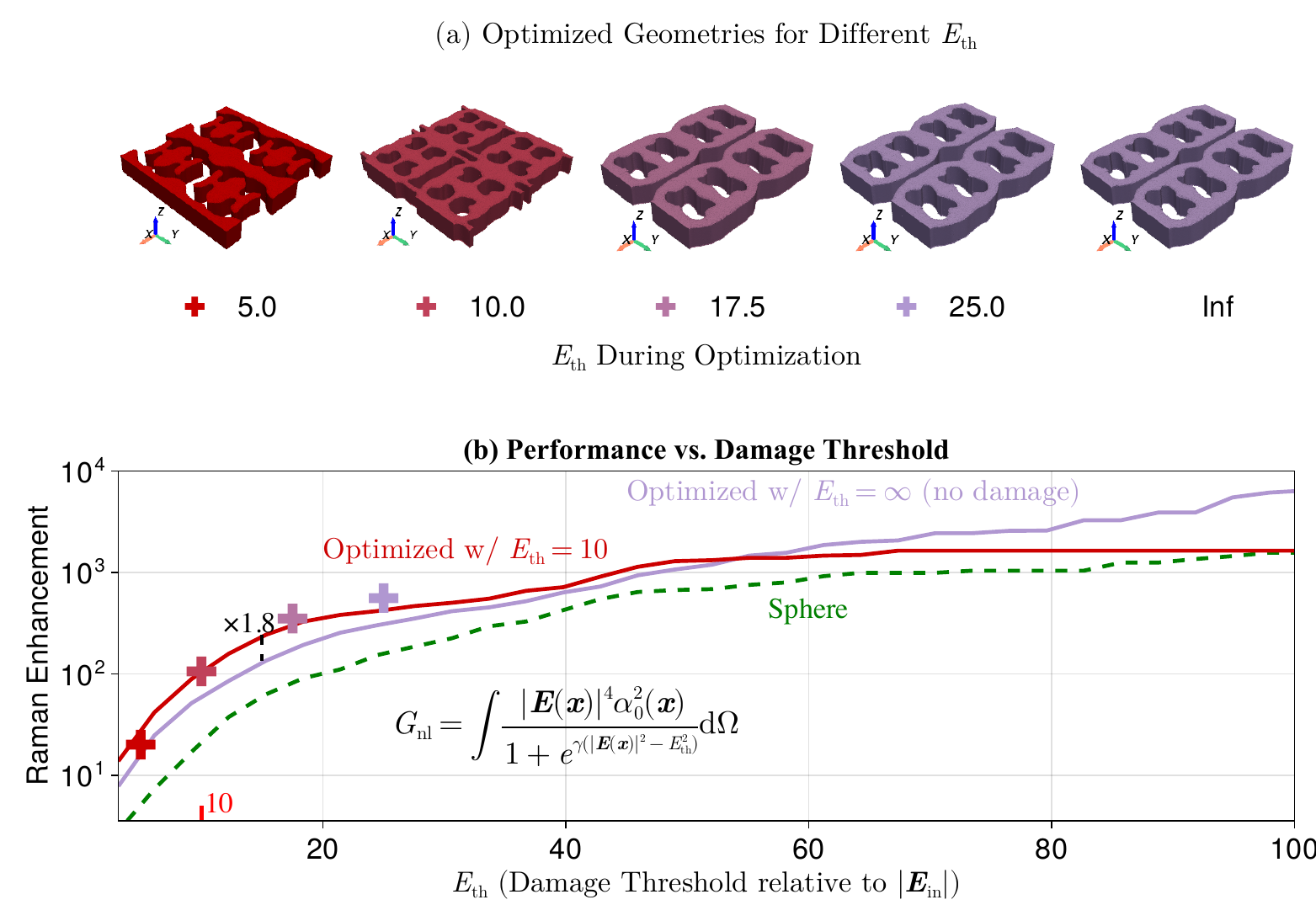} 
            \caption{Effect of nonlinear damage regularization on 3d optimized metallic structures (2d-DOFs). (a) Optimized geometries for different design thresholds $E_{\mathrm{th}}$ used during optimization. (b) Raman enhancement for structures optimized with $E_{\mathrm{th}}=10$ (red) and without any damage threshold during their optimization (purple), plotted against a varying evaluation threshold $E_{\mathrm{th}}$. Sphere array performance is shown for reference (dashed green). Markers correspond to the designs in (a) evaluated at their respective optimization $E_{\mathrm{th}}$. $G_{nl}$ is the nonlinear-damage-regularized objective (Raman enhancement penalized by quenching).}
            \label{fig:figure-5-nonlinear}
        \end{figure}

        To illustrate this adaptability, we implemented a nonlinear damage model, similar to our previous 2d~work~\cite{yao2023designing}. This model is inspired by experimental observations of molecular degradation or emission quenching in high-intensity UV SERS~\cite{yang2013ultraviolet, fang2008measurement}.  The model assumes that molecular emission is quenched if the local pump field intensity $|\vect{E}_{p}|^2$ exceeds a certain threshold intensity $E_{\mathrm{th}}^2$. This is implemented in a differentiable fashion by modifying the effective Raman polarizability term $|\alpha_0|^2$ in the objective function to $\alpha_{0,nl}^2 = |\alpha_0|^2 / (1 + e^{\gamma (|\vect{E}_\text{p}|^2 - E_\textrm{th}^2)})$, where $\gamma$ controls the sharpness of the quenching effect.
        As $E_{\mathrm{th}}$ decreases, the optimizer tends to create structures that spread the field enhancement more broadly in space to avoid exceeding the threshold, leading to entirely different topologies for low thresholds in the 2d metallic geometries, as illustrated in \figref{fig:figure-5-nonlinear}(a).

        \figref{fig:figure-5-nonlinear}(b) plots the SERS enhancement versus the evaluation threshold $E_{\mathrm{th}}$. It compares a baseline design optimized without any damage model (purple curve), an optimized sphere array (dashed green), and a design optimized with a specific threshold of $E_{\mathrm{th}}=10$ (red curve). Structures designed for lower operational thresholds tend to perform well for a wide range of damage thresholds, but eventually achieve lower peak enhancement (compared to the damage-free design) as the damage threshold is lifted. However, the improvement of the $E_{\mathrm{th}}=10$-optimized structure over the baseline at low $E_{\mathrm{th}}$ values is relatively modest (e.g., around 1.8$\times$), suggesting that the hotspots generated by the baseline metallic design (2d-DOF with $R_\mathrm{f}=20$\,nm) are already largely damped by the geometric constraints. The nonlinear objective function~$G_{nl}$, which includes the nonlinear quenching term $\alpha_{0,nl}^2$, is shown in \figref{fig:figure-5-nonlinear}(b).

    \subsection{Optimization of High-$Q$ Resonances}
        \label{sec:highq}
Dielectric SERS relies on achieving high quality factor ($Q$) resonances to obtain high field intensities (over a narrow bandwidth), as opposed to metallic SERS structures that have stronger spatial confinement and lower~$Q$ (wider bandwidth). Phenomena like bound-in-continuum (BiC) resonances theoretically allow for extremely high $Q$ factors in low-loss dielectrics~\cite{hsu2016bound}.  Recent work by Chao \textit{et al.}~\cite{chao2025sum} has shown that for infinitely periodic systems where the period $L$ equals the wavelength $\lambda$, solutions with arbitrarily large $Q$ are attainable even for lossy materials, at the expense of arbitrary vertical delocalization. In this work, we bounded the desired $Q$ to $\lesssim 500$ by including an artificial dissipation~\cite{liang2013formulation}. This limitation can, in principle, be lifted to target much higher-$Q$ dielectric structures, but that high-$Q$ regime also poses computational challenges for topology optimization.

Obtaining high-$Q$ resonances by maximizing a field response at a single frequency (such as the local density of states or our $\int |\vect{E}|^4$ objective in this paper) is known to become increasingly ``stiff'' (ill-conditioned) as $Q$ increases, causing optimization to converge more and more slowly~\cite{liang2013formulation,shaker2025eigenvalue}. However, recent work has identified a new way to overcome this obstacle by incorporating a eigenfrequency shift to keep the solver on the resonance once the $Q$ is $\gtrsim 100$~\cite{shaker2025eigenvalue}, which should be straightforward to adapt to our Raman objective. Nevertheless, we find that many low-$Q$ local optima seem to exist for periodic dielectric surfaces, so a careful initialization (perhaps with a known BiC structure) may be required to obtain an extremely high $Q$.  Vertically delocalized high-$Q$ resonances for $L \approx \lambda$~\cite{chao2025sum} also may require a large computational cell (or semi-analytical methods) to capture the long vertical tails of the resonances.

The narrow bandwidth of high-$Q$ resonances poses additional computational challenges, besides the manufacturing challenge of robustly placing a narrow resonance at the desired frequency: when an inelastic Raman shift is included, a single resonance may no longer be broad enough to enhance both the pump and emission wavelengths.  Nevertheless, in the present work we found that optimization only found singly resonant structures in \secref{sec:inelastic} even when the Raman shift was greater than $\omega_\mathrm{p}/Q$.  It is known that a sufficiently large dielectric structure can be designed to support multiple overlapping resonances, a fact that has been exploited for the design of nonlinear frequency conversion~\cite{lin2016cavity} and multi-resonant filters~\cite{chen2025inverse}, and we similarly expect multi-resonant structures to eventually become optimal for multi-resonant high-$Q$ Raman.  However, in three dimensions this increases the computational cost (by enlarging the computational cell) and further increases the number of local optima.

\section{Conclusion}
\label{sec:conclusion}
    We demonstrated a framework for 3d topology optimization of SERS substrates, maximizing spatially averaged signals from randomly oriented anisotropic emitters.
    Key to this framework is the integration of a lengthscale-constrained design paradigm for manufacturability and suppression of corner singularities, along with a novel generalization of the trace formulation~\cite{yao2022trace} to accurately account for molecular anisotropy (detailed in Appendix~A). The  regularization provided by material filtering (minimum feature size $R_\mathrm{f}$) and length-scale constraints effectively manages the tendency of the $\int|\vect{E}|^4$ periodic SERS objective to drive towards unphysical geometric singularities in 3d~\cite{andersen1978field, idemen2003confluent}. Resulting ``hotspots'' are resolved into fabricable features rather than unattainable divergences.
    The optimized structures demonstrate robustness to complex physical effects, including molecular anisotropy and inelastic Raman scattering (frequency splitting): even when evaluated with the new anisotropic figure of merit, designs optimized under an isotropic assumption stay within $\sim$20\% of those optimized with the full anisotropic model. Furthermore, the framework's adaptability was shown by incorporating a nonlinear damage model, illustrating how additional real-world constraints can be integrated into the design process.

    This research significantly advances the practical design of 3d SERS devices, paving the way for tailored substrates with enhanced sensitivity and reliability for diverse spectroscopic applications, from chemical analysis to biomedical sensing~\cite{langer2019present, pilot2019review, kneipp2007surface}. Future work will focus on experimental validation of these designs; in a specific experimental context the choice of materials and geometric constraints must be refined, along with system-specific knowledge of the distribution of Raman molecules (e.g.~our approach can accommodate a fluid layer/channel, deposition onto the surface~\cite{yao2023designing}, or nonuniform distributions~\cite{yao2022trace}).  A similar framework is also readily extensible to other distributed 3d emitter problems such as optimizing light-emitting diodes (LEDs)~\cite{janssen2010efficient}, thermal emitters~\cite{rodriguez2011frequency, luo2004thermal, kecebas2021broadband}, and scintillation detectors~\cite{roques2022framework, brenny2014quantifying}.

\appendix
\section{Derivation of Rotationally Averaged Anisotropic Correlation Term}
\label{sec:appendix-anisotropy}

It may be counter-intuitive that the rotation average of an anisotropic Raman molecule is \emph{not} mathematically equivalent to an isotropic molecule. If one could replace the true anisotropic polarizability $\alpha$ by an ``effective'' isotropic polarizability $\hat{\alpha} I$, the averaged power in \eqref{eq:trace_objective} would be proportional to $\vect{E}\vect{E}^\dagger$. In this appendix, we show that the true average contains additional tensorial structure that only reduces to the isotropic form for isotropic~$\alpha$.

\subsection{Isotropic Tensor Form}

The Raman polarizability $\alpha\in\mathbb{C}^{3\times 3}$ is a complex symmetric matrix. A molecular orientation is represented by a rotation $Q\in\text{SO}(3)$. The power expression involves the term $\alpha \vect{E} \vect{E}^\dagger \alpha^*$, which upon averaging over all orientations $Q$ yields the tensor:
\begin{equation}
\widehat{M} = \E_{Q\in\text{SO}(3)} \left[ Q\alpha Q^T \, (\vect{E}\vect{E}^\dagger)\, Q\alpha^* Q^T \right].
\label{eq:Maverage-appendix}
\end{equation}
If we assume that all orientations are equally likely (i.e.,  $Q$ is uniform in the  Haar measure on $\text{SO}(3)$~\cite{mezzadri2006generate}), then $\widehat{M}$ must be an \emph{isotropic} rank-2 tensor function of $\vect{E}$: $\widehat{M}(Q\vect{E})=Q\widehat{M}(\vect{E})Q^T$. Because the average is over \emph{real} rotations, the result must be a linear combination of the three  tensors that can be isotropically constructed from the complex vector $\vect{E}$ and its conjugate $\bar{\vect{E}}$:
\begin{equation}
\widehat{M}
  = a\,\vect{E}\vect{E}^\dagger
  + b\,\bar{\vect{E}} \vect{E}^T
  + c\,(\vect{E}^\dagger \vect{E}) I,
\label{eq:general-form}
\end{equation}
with real coefficients $a,b,c$. To determine these coefficients, we applied known integration formulations for quartic functions of $Q$~\cite{collins2003moments,weingarten1978asymptotic}, as outlined below.  Because the polarizability $\alpha$ is symmetric ($\alpha=\alpha^T$), it turns out that $b=c$. This allows us to physically regroup the terms into a Hermitian part and an anti-Hermitian ``interference'' part using only two degrees of freedom, which are convenient to express as follows:
\begin{equation}
\widehat{M} = (\alpha_{\parallel}^2 - \alpha_{\perp}^2)\,\vect{E}\vect{E}^\dagger + \alpha_{\perp}^2\!\left[(\vect{E}^\dagger \vect{E})I - \vect{E}\vect{E}^\dagger + \bar{\vect{E}}\vect{E}^T\right] \, ,
\label{eq:Mhat-structure}
\end{equation}
with $\alpha_\parallel^2 = a+2b$ and $\alpha_\perp^2 = b$. These scalars $\alpha_\parallel^2$  and $\alpha_\perp^2$ are nonnegative real coefficients determined solely by two isotropic functions of $\alpha$, the quantities $S_1 = |\Tr(\alpha)|^2$ and $S_2 = \Tr(\alpha\alpha^\dagger)$, according to the formulas 
\begin{equation}
\alpha_{\parallel}^2 = \frac{S_1 + 2S_2}{15}\qquad\text{and}\qquad\alpha_{\perp}^2 = \frac{3S_2 - S_1}{30}.
\end{equation}
We validated these results numerically by comparison to Monte-Carlo integration of  $\widehat{M}$  over $Q$.

\subsection{Properties}
The local Raman-power integrand entering Eq.~\eqref{eq:trace_objective} is of the form
\begin{align}
\vect{E}_\text{e}^\dagger \widehat{M}(\vect{E}_\text{p})\vect{E}_\text{e}
&=  (\alpha_{\parallel}^2  - \alpha_\perp^2 ) \,\big| \vect{E}_\text{e}^\dagger \vect{E}_\text{p}  \big|^2
   + \alpha_\perp^2 \,\Vert \vect{E}_\text{e} \Vert^2 \Vert \vect{E}_\text{p} \Vert^2
   - \alpha_{\perp}^2  \,\vect{E}_\text{e}^\dagger \!\left(\vect{E}_\text{p}\vect{E}_\text{p}^\dagger - \bar{\vect{E}}_\text{p}\vect{E}_\text{p}^T \right)\! \vect{E}_\text{e} \\
&= (\alpha_{\parallel}^2  - 2\alpha_\perp^2 )  \,\big| \vect{E}_\text{e}^\dagger \vect{E}_\text{p}  \big|^2
   + \alpha_\perp^2 \,\Vert \vect{E}_\text{e} \Vert^2 \Vert \vect{E}_\text{p} \Vert^2
   + \alpha_{\perp}^2 \,\big|\vect{E}_\text{e}^T \vect{E}_\text{p}\big|^2 \;\ge 0\, .
\end{align}
The invariants $S_1 = |\Tr(\alpha)|^2$ and $S_2 = \Tr(\alpha\alpha^\dagger)$ satisfy $S_1,S_2\ge 0$ by definition. Combining the expressions above, we obtain
\[
\alpha_{\parallel}^2 - \alpha_{\perp}^2
= \frac{3S_1 + S_2}{30} \ge 0,
\]
so $\alpha_\parallel^2 \ge \alpha_\perp^2$. Combined with the Cauchy--Schwarz inequality $
\big| \vect{E}_\text{e}^\dagger \vect{E}_\text{p}  \big|^2 \le  \Vert \vect{E}_\text{e} \Vert^2 \Vert \vect{E}_\text{p} \Vert^2,
$ these relations imply that the integrand is real and nonnegative for all $\vect{E}_\text{e},\vect{E}_\text{p}$, as expected.

If $\alpha$ is an isotropic tensor $\alpha_0 I$ for a scalar $\alpha_0$, then $S_1=9|\alpha_0|^2$ and $S_2=3|\alpha_0|^2$, which yields $\alpha_\perp^2 = 0$ and $\alpha_\parallel^2 = |\alpha_0|^2$, recovering the isotropic result $\widehat{M}=|\alpha_0|^2 \vect{E}\vect{E}^\dagger$. Moreover, combining Cauchy--Schwarz with the Hilbert--Schmidt inner product $(A,B)=\Tr(AB^\dagger)$ gives
\[
S_1 = |\Tr(\alpha)|^2 = |\langle\alpha,I\rangle|^2
\le \|\alpha\|_F^2\,\|I\|_F^2
= 3 S_2,
\]
with equality if and only if $\alpha$ is proportional to $I$. Since $
\alpha_\perp^2 = \frac{3S_2 - S_1}{30},
$ we have $\alpha_\perp^2 = 0$ if and only if $\alpha$ is isotropic. Thus for \emph{any} anisotropic $\alpha$, the tensor $\widehat{M}$ necessarily differs from the isotropic form.

For elastic ($\omega_\text{e}=\omega_\text{p}$) Raman scattering with matching input/output angles and polarizations, $\vect{E}_\text{e} = \vect{E}_\text{p} = \vect{E}$  and an additional simplification occurs in the integrand $\vect{E}^\dagger \widehat{M}(\vect{E})\vect{E} = (\alpha_{\parallel}^2 - \alpha_\perp^2) \Vert \vect{E}\Vert^4 + \alpha_{\perp}^2 |\vect{E}^T \vect{E}|^2$.   Even for an anisotropic $\alpha$, the Raman-power integrand can be equivalent to the isotropic form if  $|\vect{E}^T \vect{E}|=\Vert \vect{E}\Vert^2$, which is true when all components of the pump field $\vect{E}$ have the same phase (e.g.~a linearly-polarized beam in a homogeneous medium).  Thus, the anisotropic term represents a phase-dependent coupling between orthogonal polarization components.

\subsection{Determination of Coefficients}

By rotational symmetry (Eq.~\eqref{eq:general-form}), the averaged tensor must have the form
\begin{equation}
\widehat{M}(\vect{E})
  = a\,\vect{E}\vect{E}^\dagger
  + b\,\bar{\vect{E}} \vect{E}^T
  + c\,(\vect{E}^\dagger \vect{E}) I,
\label{eq:M-abc}
\end{equation}
for some real scalars $a,b,c$ depending only on $\alpha$. We now compute $a,b,c$ explicitly from the definition
\begin{equation}
\widehat{M}
= \E_{Q\in\mathrm{SO}(3)}\!\left[Q\alpha Q^T\,(\vect{E}\vect{E}^\dagger)\,Q\alpha^* Q^T\right].
\label{eq:Maverage-det}
\end{equation}

Writing $P = \vect{E}\vect{E}^\dagger$ and expanding in indices,
\begin{equation}
(\widehat{M})_{g_1 g_4}
= \sum_{g_2,g_3}\sum_{h_1,h_2,h_3,h_4}
  \alpha_{h_1 h_2}\,P_{g_2 g_3}\,(\alpha^*)_{h_3 h_4}\,
  \E_Q\!\left[Q_{g_1 h_1} Q_{g_2 h_2} Q_{g_3 h_3} Q_{g_4 h_4}\right],
\label{eq:M-g-index}
\end{equation}
where $g_1,\dots,g_4\in\{1,2,3\}$ index rows and $h_1,\dots,h_4$ index columns.

For $Q$ Haar-distributed on $\mathrm{SO}(3)$, the fourth-order moment coincides with the $\mathrm{O}(3)$ case and is given by the Collins--Śniady formula~\cite{collins2006integration,collins2003moments}:
\begin{align}
\E_Q[Q_{g_1h_1}Q_{g_2h_2}Q_{g_3h_3}Q_{g_4h_4}]
= \frac{1}{30}\Big[
&4\bigl(
   \delta_{g_1g_2}\delta_{h_1h_2}\delta_{g_3g_4}\delta_{h_3h_4}
 + \delta_{g_1g_3}\delta_{h_1h_3}\delta_{g_2g_4}\delta_{h_2h_4}
 + \delta_{g_1g_4}\delta_{h_1h_4}\delta_{g_2g_3}\delta_{h_2h_3}
 \bigr)
\label{eq:O3-moment}\\
&- \delta_{g_1g_2}\delta_{g_3g_4}
  (\delta_{h_1h_3}\delta_{h_2h_4} + \delta_{h_1h_4}\delta_{h_2h_3}) \nonumber\\
&- \delta_{g_1g_3}\delta_{g_2g_4}
  (\delta_{h_1h_2}\delta_{h_3h_4} + \delta_{h_1h_4}\delta_{h_2h_3}) \nonumber\\
&- \delta_{g_1g_4}\delta_{g_2g_3}
  (\delta_{h_1h_2}\delta_{h_3h_4} + \delta_{h_1h_3}\delta_{h_2h_4})
\Big]. \nonumber
\end{align}

Contracting the Kronecker deltas on the $h$–indices with $\alpha_{h_1 h_2}(\alpha^*)_{h_3 h_4}$ produces two rotation invariants of $\alpha$:
\begin{equation}
S_1 = |\Tr(\alpha)|^2,
\qquad
S_2 = \Tr(\alpha\alpha^\dagger),
\label{eq:S1S2-det}
\end{equation}
via
\begin{align}
\sum_{h_1,\dots,h_4} \delta_{h_1h_2}\delta_{h_3h_4}\,\alpha_{h_1h_2}(\alpha^*)_{h_3h_4}
  &= \Tr(\alpha)\Tr(\alpha^*) = S_1, \\
\sum_{h_1,\dots,h_4} \delta_{h_1h_3}\delta_{h_2h_4}\,\alpha_{h_1h_2}(\alpha^*)_{h_3h_4}
  &= \sum_{h_1,h_2} \alpha_{h_1h_2}(\alpha^*)_{h_1h_2}
   = \Tr(\alpha\alpha^\dagger) = S_2, \\
\sum_{h_1,\dots,h_4} \delta_{h_1h_4}\delta_{h_2h_3}\,\alpha_{h_1h_2}(\alpha^*)_{h_3h_4}
  &= \Tr(\alpha(\alpha^*)^T) = \Tr(\alpha\alpha^\dagger) = S_2,
\end{align}
using the symmetry $\alpha=\alpha^T$ .

Similarly, the Kronecker deltas on the $g$–indices contract $P$ into the three tensor structures
\[
P_{g_1g_4} = (\vect{E}\vect{E}^\dagger)_{g_1g_4},\quad
P_{g_4g_1} = (\vect{E}\vect{E}^\dagger)^T_{g_1g_4} = (\bar{\vect{E}}\vect{E}^T)_{g_1g_4},\quad
\delta_{g_1g_4}\,\Tr(P) = (\vect{E}^\dagger \vect{E})\,I_{g_1g_4}.
\]

Substituting Eq.~\eqref{eq:O3-moment} into Eq.~\eqref{eq:M-g-index} and collecting terms gives
\begin{align}
(\widehat{M})_{g_1g_4}
= \frac{1}{30}\Big[
   &(4S_1 - 2S_2)\,(\vect{E}\vect{E}^\dagger)_{g_1g_4}
   + (3S_2 - S_1)\,(\bar{\vect{E}}\vect{E}^T)_{g_1g_4}
   + (3S_2 - S_1)\,(\vect{E}^\dagger \vect{E})I_{g_1g_4}
\Big],
\end{align}
or, in matrix form,
\begin{equation}
\widehat{M}
= \frac{4S_1 - 2S_2}{30}\,\vect{E}\vect{E}^\dagger
 + \frac{3S_2 - S_1}{30}\,\bar{\vect{E}}\vect{E}^T
 + \frac{3S_2 - S_1}{30}\,(\vect{E}^\dagger \vect{E})I.
\label{eq:Mhat-S1S2-compact}
\end{equation}
Comparing with Eq.~\eqref{eq:M-abc}, we obtain
\begin{equation}
a = \frac{4S_1 - 2S_2}{30},
\qquad
b = c = \frac{3S_2 - S_1}{30},
\label{eq:abc-S1S2-det}
\end{equation}
so the coefficients of $\bar{\vect{E}}\vect{E}^T$ and $(\vect{E}^\dagger \vect{E})I$ are necessarily equal. 

It is convenient to re-express these in terms of
\begin{equation}
\alpha_{\parallel}^2 := \frac{S_1 + 2S_2}{15},
\qquad
\alpha_{\perp}^2 := \frac{3S_2 - S_1}{30},
\label{eq:alphas-S1S2-det}
\end{equation}
for which Eq.~\eqref{eq:Mhat-S1S2-compact} can be rearranged as
\begin{equation}
\widehat{M}
= (\alpha_{\parallel}^2 - \alpha_{\perp}^2)\,\vect{E}\vect{E}^\dagger
 + \alpha_{\perp}^2\bigl[(\vect{E}^\dagger \vect{E})I - \vect{E}\vect{E}^\dagger + \bar{\vect{E}}\vect{E}^T\bigr],
\label{eq:Mhat-final-structure-det}
\end{equation}
which is the expression used in the main text. We have also verified Eq.~\eqref{eq:Mhat-final-structure-det} numerically by Monte-Carlo integration of Eq.~\eqref{eq:Maverage-det} over random rotations $Q\in\mathrm{SO}(3)$.

\subsection{Derivation of Bounds}

Finally, we quantify the discrepancy between the true rotationally averaged objective, $g$, and the objective calculated assuming an effective isotropic polarizability, $g_{\mathrm{iso}}$. The integrated signal $g$ is a linear combination of three field-overlap scalars:
\begin{equation}
g(S_1, S_2) = c_1 \Phi_1 + c_2 \Phi_2 + c_3 \Upsilon,
\end{equation}
where $\Phi_1 = \int |\vect{E}_\mathrm{e}^\dagger \vect{E}_\mathrm{p}|^2$ (aligned overlap), $\Phi_2 = \int |\vect{E}_\mathrm{e}^T \vect{E}_\mathrm{p}|^2$ (phase-sensitive overlap), and $\Upsilon = \int |\vect{E}_\mathrm{e}|^2 |\vect{E}_\mathrm{p}|^2$ (total intensity overlap). For the standard SERS configuration (elastic scattering, single-channel matched detection with $\vect{E}_\mathrm{e} \approx \vect{E}_\mathrm{p}$), we write $\vect{E}_\mathrm{e} = \vect{E}_\mathrm{p} \equiv \vect{E}$ and obtain $\Upsilon = \Phi_1 = \int |\vect{E}|^4$. The phase-sensitive term becomes $\Phi_2 = \int |\vect{E}^T \vect{E}|^2$. By the Cauchy--Schwarz inequality ($|\vect{E}^T \vect{E}| \le |\vect{E}|^2$), this term is bounded strictly by $0 \le \Phi_2 \le \Phi_1$. Defining the field complexity parameter $t \equiv \Phi_2/\Phi_1$, we have $t \in [0,1]$.

Crucially, SERS Enhancement Factors (EF) are reported relative to a flat substrate: $\text{EF} = g^{\text{design}} / g^{\text{flat}}$. Since the exact field distribution (and thus $t$) of the flat reference is unknown \emph{a priori}, we must propagate the uncertainty using interval division ($\frac{[A, B]}{[A, B]} = [\frac{A}{B}, \frac{B}{A}]$). Comparing the isotropic approximation to the true physics yields:

\begin{itemize}
    \item \textbf{Uniaxial Molecules} ($S_1=S_2$):
    \begin{itemize}
        \item \emph{Unnormalized} ($g_{\mathrm{iso}}/g_{\mathrm{uni}}$): Isotropic assumptions overestimate by $[1.67, 2.5]\times$.
        \item \emph{Normalized EF Error} $g_{\mathrm{iso}}g^{\text{flat}}_\text{uni}/g_{\mathrm{uni}}g^{\text{flat}}_\text{iso}$: The relative error lies in $[0.67, 1.5]$.
    \end{itemize}
    \item \textbf{Traceless Molecules} ($\Tr(\alpha)=0 \implies S_1=0$) \cite{long2002raman}:
    \begin{itemize}
        \item \emph{Unnormalized} ($g_{\mathrm{iso}}/g_{\mathrm{tr}}$): Isotropic assumptions overestimate by $[2.5, 10]\times$.
        \item \emph{Normalized EF Error} $g_{\mathrm{iso}}g^{\text{flat}}_\text{uni}/g_{\mathrm{uni}}g^{\text{flat}}_\text{iso}$: The relative error lies in $[0.25, 4]$.
    \end{itemize}
\end{itemize}
These bounds confirm that optimizing for an isotropic material is a robust proxy for uniaxial molecules ($\leq 1.5\times$), and remains order-of-magnitude accurate for extreme traceless anisotropy.

\paragraph*{Funding.} {\footnotesize
This work was supported in part by the U.S. Army Research Office through the Institute for Soldier Nanotechnologies (Award No. W911NF-18-2-0048) and by a grant from the Simons Foundation through the Simons Collaboration on Extreme Wave Phenomena Based on Symmetries.
R.E.C. acknowledges support from the Danish National Research Foundation (Grant No. DNRF147 NanoPhoton).
}

\paragraph*{Acknowledgments.} {\footnotesize
We are grateful to Wenjie~Yao for helpful advice on this project.
}

\paragraph*{Disclosures.} {\footnotesize
The authors declare no conflicts of interest.
}

\paragraph*{Data availability.} {\footnotesize
The code and data used to generate the results presented in this paper are available in Ref.~\cite{DistributedEmitterOpt_jl}.
}

\bibliography{references} 

\end{document}